# On Compositional Reasoning for Guaranteeing Probabilistic Properties

Jan Olaf Blech

fortiss GmbH

**Abstract.** We present a framework to formally describe probabilistic system behavior and symbolically reason about it. In particular we aim at reasoning about possible failures and fault tolerance. We regard systems which are composed of different units: sensors, computational parts and actuators. Considering worst-case failure behavior of system components, our framework is most suited to derive reliability guarantees for composed systems. The behavior of system components is modeled using monad like constructs that serve as an abstract representation for system behavior. We introduce rules to reason about these representations and derive results like guaranteed upper bounds for system failure. Our approach is characterized by the fact that we do not just map a certain component to a failure probability, but regard distributions of error behavior and their evolvement over system runs. This serves as basis for deriving probabilities of events, in particular failure probabilities. The work presented in this paper slightly extends a complete framework and a case study which has been previously published [4]. One focus of this report is a more detailed explanation of definitions and a more comprehensive description of examples.

## 1 Introduction

The need for analysis of probabilistic systems arises in many domains connected to safety critical embedded systems. Especially the analysis of fault-tolerance establishes the need to model probabilistic distributions of failures and reason about them. Guaranteeing worst-case failure probabilities is an important prerequisite for certification of safety critical systems.

In this paper we present a framework to model probabilistic behavior of systems – especially systems' failure behavior. Our framework represents distinct parts of system behavior in an abstract monadic [21] way. We allow the modeling of behavioral entities with probabilistic distributions representing possible failures or uncertainties. When composing a system from different components, our approach allows modeling the propagation of failures through components by monadic composition of behavior associated with the components.

The second ingredient of our framework comprises rules to reason about systems. Our rules allow determining the semantic equivalence of systems and the reduction of systems to other systems such that certain properties are guaranteed to be preserved. The reduction of systems into simpler systems may be used to analyze and optimize systems.

Our approach comprises the following characteristics that all together distinguish it from existing approaches:

- Modeling of system behavior and possible faults using monad like constructs.
- Representation of uncertain/faulty behavior as distributions of possible behavior.
- Rules to reason about system behavior and distributions of values that appear in this system.

The work presented in this paper extends a complete framework and a case study which has been previously published [4]. An extended formalization and a more detailed discussion of previous formalization aspects are new to this paper. The main intended purpose of the work presented here is the usage in safety critical industrial automation systems.

### 1.1 Related Work

One long-term goal of our work is the formalization of the framework and its properties described here using a higher-order proof assistant. Early work establishing fault tolerance guarantees using theorem proving techniques is presented in [14]. Based on a formalism using extended petri nets, properties of (digital) hardware systems are shown. Furthermore, [18] describes work on guaranteeing fault-tolerance related properties using the PVS theorem prover. Here, systems and constraints are ported and proved in PVS. The presented examples come from the microprocessor and avionics domains.

Abstractions for reasoning about fault-tolerant systems in a higher-order theorem prover are presented and discussed in [22]. The abstraction aims at facilitating and standardizing the use of formal methods, especially higher-order theorem provers. Abstractions for individual message passing, faults, fault-masking, and further communication aspects are regarded.

Other early work comprises [3] which presents a Specification and Design Language (SDL) based framework. Other related work for guaranteeing fault tolerant properties using formal methods comprise the use of model-checking techniques [19] and concentrate on formal specification techniques [13]. The analysis of probabilistic system behavior is the goal of probabilistic model-checkers, like PRISM [15].

Modeling and reasoning aspects about probabilistic programs have been extensively studied in [16]. Here a language is introduced to describe probabilistic programs and reasoning about programs has a strong connection to this language. Further work on formal analysis of probabilistic systems has been done in the event-B context [10].

Like the work focused on theorem proving techniques, but unlike the (probabilistic) model checking approaches, we have a strong focus on symbolic reasoning. Unlike the existing theorem prover based work we have a strong focus on symbolically representing distributions of values and combining them. Handling errors and varying values as distributions as we do in this work allows a much richer failure analysis than assigning failure probabilities to distinct system components. For example, it allows the specification and handling of ranges in which a deviation from an optimal value is acceptable.

The framework presented in this paper builds upon work for a monadic representation of probabilities in programs [2] and an application for the analysis of cryptographic protocols [6]. Like in the analysis of cryptographic protocols, we regard possible computations that are associated with certain distributions of values. Furthermore, we regard rules that allow reasoning about sequences of such computations. In [6] and in this work the use of a monadic representation was chosen because:

- It gives a syntactic representation of the semantics of non-deterministic systems. This non-determinism can be "quantified" in the sense that different possibilities in the system execution can be assigned to different probabilities. This is achieved in combination with the use of distributions. Finally, it enables us to even specify an infinite amount of possibilities and reason about probabilities by using continuous probability distributions.
- The syntactic representation is well suited to match rules against and reason about it in a symbolic way.
- As used in [6] and as a possible future extension this gives us the possibility to reason about our system in an automatic or interactive way using, e.g., a higher-order theorem prover.

Related ideas for proving properties of (system-)processes are investigated in the context of process algebras [12,17]. The description of system behavior as a process is also featured in our framework. Nevertheless, we do not regard the presented specification and reasoning framework as a process algebra. Unlike in process algebras, our focus is not only on replacing terms by semantical equivalent ones, but rather on ensuring that a certain property is preserved during a transformation or is even improved. Like the Spi calculus [1] we target a distinct problem domain.

A future goal of our work is giving rise to certifying properties of systems in a scenario similar to [8].

As an amendment to our work on guaranteeing distinct probabilities, patterns for achieving fault tolerance have been extensively studied (e.g., [11]).

### 1.2 Overview

We present prerequisites like our monad and basic facts about probabilities in Section 2. The modeling framework is presented in Section 3. Section 4 features the rules to reason about system descriptions modeled in our framework. In Section 5 we present a larger example from the industrial control domain. In Section 6 we present a second example using discrete distributions. Finally, Section 7 gives a conclusion.

## 2 Prerequisites

In this section we describe a monad like construct to formalize computations [2,6,4]. This is needed to represent system behavior in a compositional way. The idea is to divide system behavior into different computation steps which correspond to distinct system components or computation parts. These steps realize state transitions. Traditionally, a state comprises a kind of memory, e.g., variables which are associated with values. Unlike this, in our work, we consider states in which a variable is associated with a probabilistic distribution of possible values rather than a single value. Furthermore, we present some probabilistic background knowledge.

*Distributions* Distributions may be either discrete or continuous. For a finite type $T$ the (discrete) uniformly random distribution is denoted $\$_T$. For a given value $val$ associated with a type $T$ the distribution that contains just this value (probability 100 percent) is denoted $\mathcal{U}_T(val)$. We omit the $T$ if the type is obvious from the context.

In the case of discrete distributions, the distribution can be regarded as a function that maps an element to its probability. For a given discrete distribution, the probability of an element $x$ from $D$ is denoted $D(x)$.

For the non-discrete case, $f_D$ denotes the density function of the distribution $D$. $P_D$ denotes the cumulative distribution function. Assuming a total order on the elements of $D$, the probability of all elements in $D$ that are less or equal than $x$ is denoted $P_D(x)$, In the case of normal distributions we use $\mathcal{N}(\mu, \sigma^2)$ for a normal distribution with mean $\mu$ and variance $\sigma^2$.

In mathematical distributions, the probability of all elements will sum up to 1 or in case of continuous distributions, the integral over the density function will deliver 1. In our framework, we use a slightly more general notion of distributions: in order to handle numerical approximations of distributions we allow deviations from 1.

*Monads and their composition* In our formalization, behavior is formalized using abstract computations. These are based on the *abstract computation monad* (cf. [2,6]) used for representing the changes of variable distributions. Here, we use a slightly adapted definition:

**Definition 1 (Abstract Computation).** *For a set of variables $V$ an abstract computation $M_V$ is defined by two functors* unit *and* bind.

$$\text{unit} : (V \to C_V) \to M_V \qquad \text{bind} : M_V \to \mathcal{V} \to U \to M_V$$

unit *comprises a set of initial variables distributions: $(V \to C_V)$ is a mapping from variable names of type $V$ to their distributions of type $C_V$.* bind *comprises an abstract computation, an update set of variables of type $V$: $\mathcal{V}$ and a set of updates $U$. The update set contains all variables that have to be updated. An update is a tuple:*

$$(V \times ((V \to Val_V) \to D_V))$$

*It comprises a variable to be updated of type $V$ and a function that takes a mapping from variables of type $V$ to values of type $Val_V$ and returns a distribution of type $D_V$.*

The unit constructor formalizes initial variable distributions. A bind is a single distinct computation step: The variable to distributions mapping resulting from evaluating the abstract computation (first argument of bind) is taken and for all variables in the update set (second argument of bind) the appropriate update function (in the third argument of bind) is performed resulting in new distributions for these variables.

While abstract computations are syntactical structures their semantics is given by a function [...] returning a mapping from variables to their distributions.

*Example* The semantics of the term

$$[\text{bind (bind (unit } \{(v \mapsto \$_{T_v})\}) \; \{v\} \; \{(v, v \mapsto \mathcal{U}_{T_v}(f(v)))\} \; ) \; \{v\} \; \{(v, v \mapsto \mathcal{U}_{T_v}(g(v)))\} \; ]$$

is that a value is drawn from a uniformly distributed distribution $\$_{T_v}$. A function is applied to this value, thus obtaining $f(v)$. and a distribution is made of this inner bind statement. In the next step a value is drawn from this distribution (associated with variable $v$) and a function $g$ is applied to it. The entire term again denotes a distribution. $v \mapsto D$ is used to denote a function that maps a variable $v$ to a distribution $D$.

If $f$ and $g$ are permutations the resulting distribution will be again uniformly distributed. In the remainder of this paper the "," is used to denote monadic composition of terms composed of bind and unit. Furthermore, we may omit the [...] when using the "," notation.

As a second example, we present an equation: a value is drawn from a distribution and one builds a new distribution that contains just this value. This denotes the original distribution:

$$[(v, D), (v, \mathcal{U}_{T_v}(v))] = [(v, D)]$$

In the following, we allow abusing the $=$ for denoting semantically equivalent terms, i.e., we allow omitting the [...] and may thus write:

$$(v, D), (v, \mathcal{U}_{T_v}(v)) = (v, D)$$

*Abstract Computations in Updates* Updates can contain abstract nested computations which are interpreted using [...]. These abstract computations may work on a different variable domain. Furthermore, only the variable of the update is effected outside such a nested abstract computations. This formalism may be used to realize abstract scopes of variables. We omit the [...] when using the "," notation.

The following shows an example:

$$(v, ((v, D_v), (v', D_{v'})))$$

The variable $v$ is updated in a nested abstract computation with the distribution $D_v$. The update to $v'$ is not exported out of this scope and $v'$ may not even be defined outside this inner scope.

*Events and Probabilities* We define probabilistic *events* on abstract computations. Events take values associated with variables $val_1, ..., val_n$ and return a truth value:

**Definition 2 (Event).** *An event E is a function:*
$$E(val_{v_1}, ..., val_{v_n}) \rightarrow \{true, false\}$$
*with $v_1, ..., v_n \in V$ and $val_{v_1} \in T_{v_1}, ..., val_{v_n} \in T_{v_n}$.*

An Event $E$ can be applied to an abstract computation $C$ thereby specifying a value between 0 and 1 stating the probability that $E$ does hold after the computation of $C$. We denote this: $Pr([C]E) \in [0, 1]$.

*Example* An event that states that a value drawn from a boolean distribution associated with a variable $b$ is true is formalized as $b = true$. The probability of this event for a uniform boolean distribution:

$$Pr([(b, \$_{\{true, false\}})](b = true)) = 0.5$$

# 3 Our Modeling Framework

In this section we describe the framework for specification of systems using abstract computations from Section 2.

## 3.1 Monadic Representation of Probabilistic Systems

Complex systems composed of different components can be described with abstract computations. The components are modeled as functions that take some values bound to distinct variables and map them to their distributions.

Different components realizing a certain behavior may be modeled independently. They may be composed so that a system description can be realized. In the industrial automation domain – which we focus on – typical components for modeling systems comprise sensors, actuators, and plain computations. Sensors and actuators comprise a distinct failure behavior which can be modeled by using distributions. Computations may also be formalized in a way that they are associated with a failure behavior. This can be used, e.g., to model potential failures in the underlying hardware.

We give a small overview on modeling system behavior using our formalism.

*Sequential Composition* Given two system components $A$ and $B$ performing a certain task and formalized using our abstract computation. To activate $A$ and $B$ sequentially we may combine them using standard monadic composition:

$$A\ ,\ B$$

$A$ gets executed before $B$ and all values computed in $A$ are accessible in $B$.

*Parallel Composition* Two components $A$, $B$ can be combined in parallel composition. This is denoted:

$$\begin{pmatrix} A \\ B \end{pmatrix}$$

$A$ and $B$ should not depend on each other, i.e., no write access to a value which is used by the other component. Thus, it is possible to linearize parallel composition to:

$$A\ ,\ B \quad \text{or} \quad B\ ,\ A$$

*Conditional Structures* Conditional structures depending on an expression $e$ can be realized in a straightforward way:

$$if\ \ e\ \ then\ \ A_1, ..., A_n\ \ else\ \ B_1, ...B_n$$

Distributions of variables' values can be effected in different ways by different branches of conditional expressions. For this reason, rather complex definitions of distributions may occur due to conditional structures.

*Looping Structures* Given a loop body $A$, looping structures may be formalized in the following way, resembling process algebras:

$$L := A , L$$

Our framework bears some similarity to the way one describes Programmable Logic Controllers (PLCs) by, e.g., using the IEC-61131–3 standard [20]. It is intended to be an extension of the property certification proposed in [7] for PLCs.

### 3.2 An Example

Here we present a generic example that realizes a composed component made from sensors and a computation.

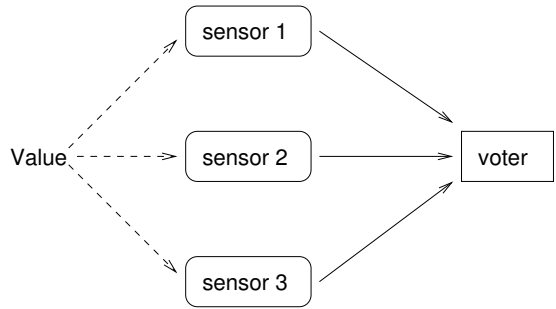

Fig. 1: Voting Example

The example system in Figure 1 realizes a voting. This is a fault tolerance mechanism that aims at eliminating errors occurring in individual components by replicating them. In our case a sensor is replicated. One physical value $x$ is read by three different sensors. Each of these sensors may read a wrong value: some noise is added to $x$ which corresponds to the distribution $\mathcal{N}_E$ and is associated with the variable names $e_1$, $e_2$ and $e_3$. Thus, this noise is independent for each sensor. The voting reads the three sensor values $v_1$, $v_2$, $v_3$ and can, e.g., be realized by computing the arithmetic mean $r$.

$voter\_mean(x) \equiv$

$$\begin{pmatrix} e_1, \mathcal{N}_E \\ e_2, \mathcal{N}_E \\ e_3, \mathcal{N}_E \end{pmatrix} , \begin{pmatrix} v_1, \mathcal{U}_T(x+e_1) \\ v_2, \mathcal{U}_T(x+e_2) \\ v_3, \mathcal{U}_T(x+e_3) \end{pmatrix} , (r, \mathcal{U}_{T_r} \left( \tfrac{v_1+v_2+v_3}{3} \right))$$

An alternative realization – in the case of discrete noise given by the distribution $E$ – is given below:

$$voter\_2(x) \equiv$$

$$\begin{pmatrix} e_1, E \\ e_2, E \\ e_3, E \end{pmatrix}, \begin{pmatrix} v_1, \mathcal{U}_T(x+e_1) \\ v_2, \mathcal{U}_T(x+e_2) \\ v_3, \mathcal{U}_T(x+e_3) \end{pmatrix}, (r, \mathcal{U}_{T_r} \begin{pmatrix} if & v_1 = v_2 & then & v_1 \\ if & v_1 = v_3 & then & v_1 \\ if & v_2 = v_3 & then & v_2 \\ & & else & v_3 \end{pmatrix})$$

Here we compare the values from the three different sensors. If two are equal we return one of these equal values. If all three are different, we return the value of the third sensor.

## 4 Deductive Rules

We have established rules to reason about system descriptions which are based on abstract computations. Two goals can be distinguished:

- Rules that transform system descriptions into other system descriptions. A special case of transformation rules perform semantically equivalent transformations.
- Rules that allow reasoning about probabilities of certain events. Some of these rules allow the transformation of systems with respect to certain events.

Furthermore, we can distinguish rules that work on discrete and those that work on continuous distributions.

### 4.1 Soundness and Semantics of Rules

Our abstract computations represent both: system components and a semantical representation of them. Further elements that carry semantical meaning are the application of an event to an abstract computation and the probability function $Pr$. For this reason, proving soundness of our rules to reason about abstract computations does not need to take a transformation between syntax and semantics into account. In order to prove soundness one proves that certain semantical aspects are preserved during a rule application. In particular our rules are proven sound with respect to the following notions of correctness:

- Rules that transform system descriptions into other system descriptions: These rules have the following form:

$$\dfrac{AbstractComputation(val_1, ..., val_n) \quad\quad \text{additional assumptions } (val_1, ..., val_n, val'_1, ..., val'_n)}{AbstractComputation'(val'_1, ..., val'_n)} \lesssim$$

Soundness of such a rule is established by proving the following lemma:

$$\forall\ E\ val_1\ ...\ val_n\ val'_1\ ...\ val'_n\ \epsilon.$$
$$\text{additional assumptions } (val_1, ..., val_n, val'_1, ..., val'_n) \wedge$$
$$Pr([AbstractComputation(val_1, ..., val_n)]E) = \epsilon$$
$$\longrightarrow$$
$$Pr([AbstractComputation'(val'_1, ..., val'_n)]E) = \epsilon$$

Thus, our notion of correctness states that the probability of all possible events is preserved while transforming an abstract computation. We use $\lesssim$ in this paper to denote probability preservation for all possible events.
- Rules that allow reasoning about probabilities of certain events directly correspond to a lemma stating a fact on abstract computations.

Rules are applied by matching the bottom part of the rule to a system description and event. The application reduces the expression to the upper part of the rule.

### 4.2 Basic Rules

Here we present basic rules to handle system descriptions within our framework. Their soundness is proved in our scheme by using the properties of abstract computations.

*Function Propagation Rule* We have established rules to reason and simplify our monadic system descriptions. Simple rules comprise, e.g., function propagation.

$$\frac{\forall i \leq m, x'_i \neq x' \qquad (x'', \mathcal{U}_T(g(f(x_1, ..., x_n), x'_1, ...x'_m)))}{(x'', ..., (x', \mathcal{U}_T(f(x_1, ..., x_n))) , (x'', \mathcal{U}_T(g(x', x'_1, ..., x'_m))))} \lesssim$$

A generalized version of this rule does not require the usage of nested abstract computations. It is sufficient that $x'$ is not used again in later expressions.

*Omitting Unused Parts Rule* Unused parts of an expression may be omitted.

$$\frac{\forall i \leq n, x'_i \neq x}{(x, X'(x'_1, ..., x'_n))} \lesssim$$
$$\frac{(x, X'(x_1, ..., x_n)) , (x, X'(x'_1, ..., x'_n))}{} \lesssim$$

This rule performs a kind of dead-code elimination

*Congruence Exchange Rule* Semantical equivalent parts may be replaced by each other.

$$\frac{A, B, C \qquad B \lesssim B'}{A, B', C} \lesssim$$

*Permutation Rule* Parts may be permuted if they do not depend on each other.

$$\forall i \leq n, x_i \neq x'$$
$$\forall i' \leq m, x'_i \neq x$$
$$\frac{(x, X(x_1, ..., x_n)) , (x', X'(x'_1, ..., x'_m))}{(x', X'(x'_1, ..., x'_m)) , (x, X(x_1, ..., x_n))} \lesssim$$

### 4.3 General Rules Relating Events and Probabilities

The following rules bridge the gaps between abstract computations, events and probabilities. Soundness is established by using the definitions from Section 2.

*Event Approximation for Continuous Distributions Rule* The following rule allows the numerical approximation of a probability. Given the continuous distribution $D$, its probability density function $f_D$ and an order on the elements with this distribution $\leq$. We can use an approximation of $f_D$ : $f_A$ in order to guarantee a certain maximal probability of an event that checks whether a certain value drawn from $D$ is below some upper bound $a$. $P_D$ and $P_A$ denote the cumulative probability functions (or approximation in case of $P_A$) that correspond to $f_D$ and $f_A$.

$$\frac{\forall v\, .f_D(v) \leq f_A(v) \qquad P_A(a) < \epsilon}{Pr([(x,D)](x \leq a)) < \epsilon}$$

We can use this rule to simplify expressions and leave the subgoal $P_A(a) < \epsilon$ for numerical approximation. Similarly the following rule holds:

$$\frac{\forall v\, .f_A(v) \leq f_D(v) \qquad 1 - P_A(a) < \epsilon}{Pr([(x,D)](x \geq a)) < \epsilon}$$

Note that for the first rule the approximation $\int_{-\infty}^{\infty} f_A(v)\, dv$ will be greater or equal than 1 while in the second rule it while be smaller or equal than one. This is ensured by the first condition in both rules.

*Range Event Splitting Rule* The following rule may be used to split an event stating that a variable is outside a certain range into two independent subevents.

$$\frac{Pr([(x,D)](x \geq a)) < \epsilon_1 \qquad Pr([(x,D)](x \leq b)) < \epsilon_2}{Pr([(x,D)](x \geq a \vee x \leq b)) < \epsilon_1 + \epsilon_2}$$

### 4.4 Rules for Normal Distributions

Here we present a few rules valid for normal distributions. Their soundness can be easily proven since they correspond to well known facts about normal distributions.

*Normal Distribution Rule* We have established a rule for combining values originating from different normal distributions $\mathcal{N}(\mu_i, \sigma_i^2)$.

$$\frac{(x', \mathcal{N}(\mu_1 + ... + \mu_n, \sigma_1^2 + ... + \sigma_n^2))}{(x', \begin{pmatrix} x_1, \mathcal{N}(\mu_1, \sigma_1^2) \\ ... \\ x_n, \mathcal{N}(\mu_n, \sigma_n^2) \end{pmatrix}, (x', \mathcal{U}_T(x_1 + ... + x_n)))} \lesssim$$

*Normal Distribution Probability Event Rule* Another rule relates normal distributions, events, and probabilities.

$$\frac{Pr([(x,\mathcal{N}(\mu,\sigma^2))](x \leq \mu - a)) < \epsilon \qquad a \leq \sigma \qquad \sigma \leq \sigma'}{Pr([(x,\mathcal{N}(\mu,\sigma'^2))](x \leq \mu - a)) < \epsilon}$$

It corresponds to standard facts on normal distributions. Likewise the following rule holds.

$$\frac{Pr([(x,\mathcal{N}(\mu,\sigma^2))](x \geq \mu + a)) < \epsilon \qquad a \leq \sigma \qquad \sigma \leq \sigma'}{Pr([(x,\mathcal{N}(\mu,\sigma'^2))](x \leq \mu + a)) < \epsilon}$$

These rules relate abstract computations with typical events that check whether a normally distributed variable has a value in a certain range. Their correctness is established by looking at the probability density function for the normal distribution,

$$P_\mathcal{N}(x) = \tfrac{1}{\sigma\sqrt{2\pi}} exp\left(-\tfrac{1}{2}\left(\tfrac{x-\mu}{\sigma}\right)^2\right)$$

its derivation and especially the points $\mu - \sigma$ and $\mu + \sigma$.

*Voting Abstraction Rule* A specialized rule for simplifying the semantics of voting can be established. Setups containing a voting computing the mean of several values which may be influenced by $\mathcal{N}$ distributed errors can be simplified by using the following rule:

$$\frac{(r,((e,\mathcal{N}(\tfrac{\mu_1+...+\mu_n}{n},\tfrac{\sigma_1^2+...+\sigma_n^2}{n^2})),(r,\mathcal{U}_{T_r}(x+e))))}{(r,\begin{pmatrix}e_1,\mathcal{N}(\mu_1,\sigma_1^2)\\...\\e_n,\mathcal{N}(\mu_n,\sigma_n^2)\end{pmatrix},\begin{pmatrix}v_1,\mathcal{U}_T(x+e_1)\\...\\v_n,\mathcal{U}_T(x+e_n)\end{pmatrix},(r,\mathcal{U}_{T_r}\left(\tfrac{v_1+...+v_n}{n}\right)))} \lesssim$$

The soundness of this rule is derived from the Normal Distribution Rule and the Function Propagation Rule.

### 4.5 Rules for Discrete Distributions

Some facts can only be derived for discrete distributions.

*Discrete Probability Computation Rule* The following rule may be used to compute probabilities of expressions containing discrete distributions:

$$\frac{\sum_{x' \in D} D(x') \cdot Pr([(x,F(x))]E)}{Pr([(x',D),(x,F(x')]E)}$$

A distribution is eliminated by considering the probabilities of all possible values independently.

*Event Probability Weakening Computation Rule* This rule relates events and probabilities of discrete distributions.

$$\frac{Pr([(x,D')](x = y)) < \epsilon \qquad D(y) \leq D'(y)}{Pr([(x,D)](x = y)) < \epsilon}$$

It is to be used with distinct events that check for the appearance of certain values. It can also be used to approximate distributions similar to the Event Approximation for Continuous Distributions Rule.

*Additional Rules and Approximations* Our presented rules may be used to perform simplifications in order to discover a certain correctness result. Additionally, after these simplifications we may use numerical methods, e.g., to approximate probabilities of expressions that are not easily handled in an algebraic way.

## 5 Continuous Case Study

In this case study we regard a work piece on a conveyor belt. Actuators and sensors are used to bring it close to a desired position. A version of the voting element (cf. Sec-

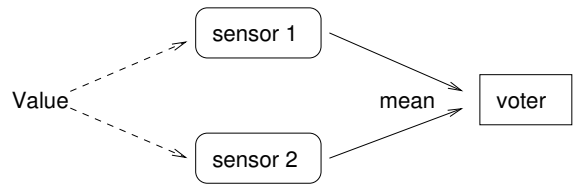

Fig. 2: Voting element

tion 3.2) with two sensors depicted in Figure 2 is used in our case study:

$$vote2(x) \equiv$$
$$\begin{pmatrix} e_1, \mathcal{N}(\mu_E, \sigma_E^2(x)) \\ e_2, \mathcal{N}(\mu_E, \sigma_E^2(x)) \end{pmatrix} \ , \ \begin{pmatrix} v_1, \mathcal{U}_{T_{v_1}}(x + e_1) \\ v_2, \mathcal{U}_{T_{v_2}}(x + e_2) \end{pmatrix} \ , \ (r, \mathcal{U}_{T_r}\left(\tfrac{v_1+v_2}{2}\right))$$

The variance is parameterized with the current position $x$ since the sensors accuracy depends on the current position of the work piece. Since both sensors behave in the same way, our rules to simplify normal distributions are still applicable.

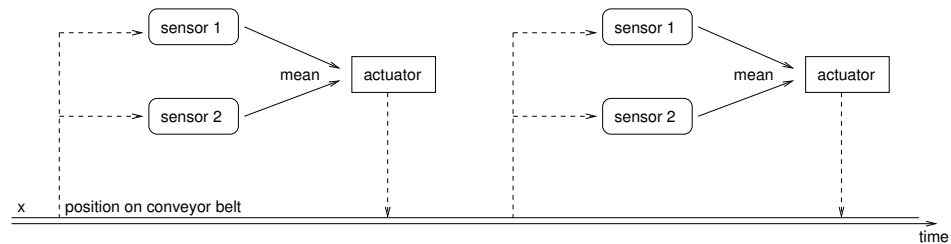

Fig. 3: Conveyor belt control

In the actual case study a work piece is put on a conveyor belt. Our goal is to bring this work piece close to a position $p$. We can measure the position $x$ on the conveyor belt by using two sensors. The sensors can be influenced by normal distributed errors. We can activate the conveyor belt to adjust the position. This repositioning will also be influenced by some possible error.

Our conveyor belt case study is shown in Figure 3. In our formalism it is described as shown in Figure 4. The following sequence is performed two times in order to achieve a good positioning of the work piece:

– We read the value of the work piece via two different sensors. The value of the position of the work piece $x$ may be effected by $\mathcal{N}(\mu_E, \sigma_E(x))$ distributed sensor errors $e_1$, $e_2$ while reading them with the sensors. The sensors write the values which they have read to variables $v_1$, $v_2$. We perform a mean voting on their results and store it in a variable $r$.
– We activate the conveyor belt in order to move the work piece close to $p$ by giving it $p - r$ as repositioning information.
– The conveyor belt is not necessarily moved exactly by the requested distance, but again, the actuator introduces an error which is $\mathcal{N}(\sigma'_E, \sigma'_E)$ distributed.

$conv\_belt(x, p) \equiv$

$(r, \begin{pmatrix} e_1, \mathcal{N}(\mu_E, \sigma_E^2(x)) \\ e_2, \mathcal{N}(\mu_E, \sigma_E^2(x)) \end{pmatrix}, \begin{pmatrix} v_1, \mathcal{U}_{T_{v_1}}(x + e_1) \\ v_2, \mathcal{U}_{T_{v_2}}(x + e_2) \end{pmatrix}, (r, \mathcal{U}_{T_r}(\frac{v_1 + v_2}{2}))),$

$(p', \mathcal{U}_T(p - r)), (e, \mathcal{N}(\mu'_E, \sigma_E'^2)), (x, \mathcal{U}_T(p' \cdot (1 + e))),$

$(r, \begin{pmatrix} e_1, \mathcal{N}(\mu_E, \sigma_E^2(x)) \\ e_2, \mathcal{N}(\mu_E, \sigma_E^2(x)) \end{pmatrix}, \begin{pmatrix} v_1, \mathcal{U}_{T_{v_1}}(x + e_1) \\ v_2, \mathcal{U}_{T_{v_2}}(x + e_2) \end{pmatrix}, (r, \mathcal{U}_{T_r}(\frac{v_1 + v_2}{2}))),$

$(p', \mathcal{U}_T(p - r)), (e, \mathcal{N}(\mu'_E, \sigma_E'^2)), (x, \mathcal{U}_T(p' \cdot (1 + e)))$

Fig. 4: Conveyor belt control (formal description)

Our goal is to ensure that the conveyor belt will be in a close range to a distinct position $p$ with a certain probability. We are interested in questions like: what is the probability that the difference between optimal and actual position is smaller than a given constant $l$: $p - x < l$:

$$Pr([conv\_belt(x, p)](l \leq p - x)) < \epsilon$$

The maximal probability that this is not the case shall be $\epsilon$.

In order to do so, we apply our rules defined in Section 4 to the system definition. The goal is to derive the distribution of $p$ once the system has reached its terminal state. In a first step, we simplify the voting of the sensors (Congruence Exchange, Voting Abstraction, Permutation, associativity of monads). Thus, we derive the following simplified system description:

$(r, \mathcal{N}(x + \mu_E, \frac{\sigma_E^2(x)}{2})), (p', \mathcal{U}_T(p - r)), (e, \mathcal{N}(\mu'_E, \sigma_E'^2)), (x, \mathcal{U}_T(p' \cdot (1 + e))),$

$(r, \mathcal{N}(x + \mu_E, \frac{\sigma_E^2(x)}{2})) , (p', \mathcal{U}_T(p - r)) , (e, \mathcal{N}(\mu'_E, \sigma'^2_E)) , (x, \mathcal{U}_T(p' \cdot (1 + e)))$

This can be further simplified (Congruence Exchange, Function Propagation):
$(r, \mathcal{N}(x + \mu_E, \frac{\sigma_E^2(x)}{2})) , (e, \mathcal{N}(\mu'_E, \sigma'^2_E)) , (x, \mathcal{U}_T((p - r) \cdot (1 + e))) ,$

$(r, \mathcal{N}(x + \mu_E, \frac{\sigma_E^2(x)}{2})) , (e, \mathcal{N}(\mu'_E, \sigma'^2_E)) , (x, \mathcal{U}_T((p - r) \cdot (1 + e)))$

We do not touch the expression containing the product of normal distributed variables.

Since the initial position $x$ and $p$ are known, at this stage one can numerically handle the expression in order to convince oneself that numerical constraints on these distributions are met. Computer algebra systems allow for an over-approximation of the distribution above. Thus, one can apply the Event Approximation for Continuous Distributions Rule and ensure that the given failure probability $\epsilon$ is met.

Furthermore, given the expression above one can consider possible optimization alternatives, like, e.g., updating the sensors or actuators so that they feature a better error distribution. Another optimization possibility would be to replicate the sensor-actuator part another time. One can numerically recalculate the the results and convince oneself that they meet a certain $\epsilon$.

## 6  A Discrete Case Study

Figure 5 shows a second case study. Again it is taken from the area of industrial automation. A sensor decides whether a work piece is red or blue. Depending on the sensor choice the work piece is sorted to one of the two stacks: $stack1$ for red workpieces, $stack2$ for blue workpieces. Its formal description is given in Figure 6. The sensor can return an erroneous result: It can return blue for a red working piece and vice versa. Likewise the actuator may sort a workpiece to a wrong stack. The probabilities of such a malfunctioning of sensor and actor are formalized using the distributions $D_{red}$, $D_{blue}$, $D_{stack1}$, and $D_{stack2}$.

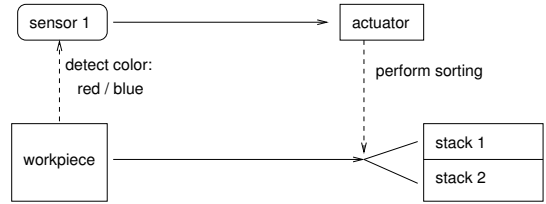

Fig. 5: Discrete sensor and actuator

Consider the following example distribution density functions:
$f_{red} = \lambda x. \begin{pmatrix} if & x = red & then & 0.95 \\ if & x = blue & then & 0.05 \end{pmatrix}$

$$DiscreteSort(val_c) \equiv$$

$$(c, \mathcal{U}_{T_c}(val_c)),$$
$$\left(c', \begin{pmatrix} if & c = red & then & D_{red} \\ if & c = blue & then & D_{blue} \end{pmatrix}\right),$$
$$\left(s, \begin{pmatrix} if & c' = red & then & D_{stack1} \\ if & c' = blue & then & D_{stack2} \end{pmatrix}\right)$$

Fig. 6: Discrete sensor and actuator(formal description)

$$f_{blue} = \lambda x. \begin{pmatrix} if & x = red & then & 0.05 \\ if & x = blue & then & 0.95 \end{pmatrix}$$
$$f_{stack1} = \lambda x. \begin{pmatrix} if & x = stack1 & then & 0.95 \\ if & x = stack2 & then & 0.05 \end{pmatrix}$$
$$f_{stack2} = \lambda x. \begin{pmatrix} if & x = stack1 & then & 0.01 \\ if & x = stack2 & then & 0.99 \end{pmatrix}$$

With these definitions we can compute the probability that a red workpiece will be put on the wrong stack by using the Discrete Probability Computation Rule. We do this by stating the following expression using an appropriate event:

$$Pr([DiscreteSort(red)](s = stack2))$$

The expression can be simplified by using the Discrete Probability Computation Rule. Finally we get the following probability:

$$1 - (0.95 \cdot 0.95 + 0.05 \cdot 0.01) = 0.0970$$

Accordingly we can state an event stating that a blue workpiece will be put on the wrong stack:

$$Pr([DiscreteSort(blue)](s = stack1))$$

Simplification gives us the following probability:

$$1 - (0.95 \cdot 0.99 + 0.05 \cdot 0.05) = 0.057$$

## 7 Conclusion

We propose a framework to specify systems and their behavior and possible divergences that might occur in these systems. We present some extensions and more details compared to a previous version of this paper [4]. Our framework is build upon a monadic representation of execution steps which allows the representation of possible errors and their probabilities. A rule based logic is presented to reason about our system description and perform algebraic simplifications. Ultimately we derive guarantees for the system description. Our rules encapsulate common tasks, reasoning is not limited to these rules. In particular we allow numerical approximations. We present two case studies to demonstrate possible usage scenarios of our framework.

*Future Work* As a long term goal, our framework is intended to be used with the Coq [9] theorem prover. Depending on the usage scenario, an implementation in another verification / analysis tool might also be an option. We continue looking at additional case studies.

*Acknowledgment* This work has been supported by the European research project ACROSS under the Grant Agreement ARTEMIS-2009-1-100208.

## References

1. M. Abadi and A. D. Gordon. A calculus for cryptographic protocols: The spi calculus. 4th ACM Conference on Computer and Communications Security, ACM Press, 1997.
2. P. Audebaud and C. Paulin-Mohring. Proofs of randomized algorithms in Coq. Science of Computer Programming. 2008.
3. S. Ayache, E. Conquet, P. Humbert, C. Rodriguez, J. Sifakis, and R. Gerlich. Formal methods for the validation of fault tolerance in autonomous spacecraft. International Symposium on Fault-Tolerant Computing. 1996. (FTCS '96)
4. J. O. Blech. Probabilistic Compositional Reasoning for Guaranteeing Fault Tolerance Properties.
5. 5th International Conference On Principles Of Distributed Systems, vol. 7109 of LNCS, Springer, 2011.
6. J. O. Blech. Proving the Security of ElGamal Encryption Via Indistinguishability Logic. ACM Symposium On Applied Computing. 2011.
7. J. O. Blech, A. Hattendorf, and J. Huang. An Invariant Preserving Transformation for PLC Models. IEEE International Workshop on Model-Based Engineering for Real-Time Embedded Systems Design, 2011.
8. J. O. Blech and M. Périn. Generating Invariant-based Certificates for Embedded Systems. In *ACM Transactions on Embedded Computing Systems* (TECS). *to appear*
9. The Coq development team: The Coq Proof Assistant Reference Manual v8.3 (2010) Available at http://coq.inria.fr.
10. S. Hallerstede and T. S. Hoang Qualitative Probabilistic Modelling in Event-B*. Integrated Formal Methods. vol. 4591 of LNCS, Springer-Verlag, 2007.
11. Robert Hanmer. Patterns for Fault Tolerant Software. Wiley, ISBN: 978-0-470-31979-6, October 2007.
12. C. A. R. Hoare. Communicating Sequential Processes. Prentice Hall, ISBN 0-13-153289-8, 1985.
13. R. Jeffords, C. Heitmeyer, M. Archer and E. Leonard. A Formal Method for Developing Provably Correct Fault-Tolerant Systems Using Partial Refinement and Composition. Formal Methods 2009, vol. 5850 of LNCS, Springer-Verlag, 2009.
14. J. Kljaich, B. T. Smith, A. S. Wojcik. Formal Verification of Fault Tolerance Using Theorem-Proving Techniques. IEEE Transactions on Computers. Volume 38 Issue 3, March 1989.
15. M. Kwiatkowska, G. Norman and D. Parker. PRISM: Probabilistic Symbolic Model Checker. vol. 2324 of LNCS, Springer-Verlag, 2002.
16. A. McIver and C. Morgan. Abstraction, Refinement and Proof for Probabilistic Systems. Springer-Verlag, 2005.
17. R. Milner. Communicating and Mobile Systems: the Pi-Calculus. Springer Verlag, ISBN 0-521-65869-1, 1999.

18. S. Owre, J. Rushby, N. Shankar, F. von Henke. Formal verification for fault-tolerant architectures: prolegomena to the design of PVS. IEEE Transactions on Software Engineering, Feb 1995.
19. W. Steiner, J. Rushby, M. Sorea, and H. Pfeifer. Model Checking a Fault-Tolerant Startup Algorithm: From Design Exploration To Exhaustive Fault Simulation. The International Conference on Dependable Systems and Networks. IEEE Computer Society, 2004.
20. Programmable controllers - Part 3: Programming languages, IEC 61131-3: 1993, International Electrotechnical Commission, 1993.
21. Philip Wadler. The essence of functional programming. 19'th Symposium on Principles of Programming Languages, ACM Press, January 1992.
22. L. Pike, J. Maddalon, P. Miner and A. Geser. Abstractions for Fault-Tolerant Distributed System Verification. Theorem Proving in Higher-Order Logics, vol 3223 of LNCS, Springer-Verlag, 2004.